\documentclass[traditabstract]{aa} 
\usepackage{graphicx}
\usepackage{txfonts}
\usepackage{natbib}
\usepackage{rotating}

\newcommand{\ergs}{erg\ s$^{-1}$}
\newcommand{\kms}{km\ s$^{-1}$}

\newcommand{\xmm}{{\sc XMM}\emph{-Newton}}
\newcommand{\ros}{\emph{{\sc ROSAT}}}
\newcommand{\hd}{HD\,155806}
\newcommand{\ha}{H$\alpha$}

\newcommand{\lxlb}{$L_{\rm X}/L_{\rm BOL}$}
\newcommand{\loglxlb}{$\log[L_{\rm X}/L_{\rm BOL}]$}

\begin{document}
   \title{Hot and cool: two emission-line stars with constrasting behaviours in the same \xmm\ field\thanks{Based on observations collected with XMM-Newton, an ESA Science Mission with instruments and contributions directly funded by ESA Member States and the USA (NASA).}}

   \author{Ya\"el Naz\'e
          \inst{1}\fnmsep\thanks{Research Associate FNRS} \and Gregor Rauw\inst{1}\fnmsep$^{\star\star}$ \and Asif ud-Doula\inst{2}
          }

   \institute{GAPHE, D\'epartement AGO, Universit\'e de Li\`ege, All\'ee du 6 Ao\^ut 17, Bat. B5C, B4000-Li\`ege, Belgium\\
              \email{naze@astro.ulg.ac.be}
   \and Department of Physics, Morrisville State College, Morrisville, NY 13408, USA
             }

   \titlerunning{The X-ray emission of \hd\ and 1RXS\,J171502.4$-$333344}

 
  \abstract
{
High-energy emissions are good indicators of peculiar behaviours in stars. We have therefore obtained an \xmm\ observation of \hd\ and 1RXS\,J171502.4$-$333344, and derived their spectral properties for the first time.
The X-ray spectrum of \hd\ appears soft, even slightly softer than usual for O-type stars (as shown by a comparison with the O9 star HD\,155889 in the same \xmm\ field). It is well-fitted with a two-component thermal model with low temperatures (0.2 and 0.6 keV), and it shows no overluminosity (\loglxlb=$-$6.75). The high-resolution spectrum, though noisy, reveals a few broad, symmetric X-ray lines ($FWHM\sim2500$\,\,\kms). The X-ray emission is compatible with the wind-shock model and therefore appears unaffected by the putative dense equatorial regions at the origin of the Oe classification. 
1RXS\,J171502.4$-$333344 is a nearby flaring source of moderate X-ray luminosity (\loglxlb=$-$3), with a soft thermal spectrum composed of narrow lines and presenting a larger abundance of elements (e.g. Ne) with a high first ionization potential (FIP) compared to lower-FIP elements. All the evidence indicates a coronal origin for the X-ray emission, in agreement with the dMe classification of this source.
}
   \keywords{X-rays: stars -- X-rays: individual: 1RXS\,J171502.4$-$333344 -- Stars: coronae -- Stars: individual: HD\,155806 -- Stars: individual: HD\,155889 -- Stars: emission-line }

   \maketitle
%

\section{Introduction}

Over the past decades, stars of almost all spectral types have been identified as X-ray emitters \citep[for a review see][]{gud09}. However, the physical processes at the origin of this high-energy emission differ throughout the HR diagram. Coronae play a large role in the X-ray emission of F to M stars, as well as in young T Tauri objects. For these sources, flares, narrow X-ray lines, tight relation between rotation and X-ray luminosity, as well as changing abundance patterns, were detected, indicating that magnetic processes are heating up the X-ray plasma. In contrast, the X-ray emission from early-type stars appears highly dependent on the stellar winds. The high energies thus uniquely probe the mass-loss processes, and additional emission is easily spotted when two winds collide in binaries. Peculiar stellar phenomena (e.g. accretion/decretion discs, jets, magnetically-confined winds) are also shown to affect the X-ray emission. The X-ray domain is therefore a unique tool for analysing the physical processes at work in the envelopes of stars.

This paper provides the first X-ray investigation of two peculiar emission-line stars at two opposite extremes of the HR diagram: the hot Oe star \hd\ and the cool dMe object 1RXS\,J171502.4$-$3333440. Section 2 presents our dataset, Sects. 3 and 4 give the observed X-ray properties of the Oe star \hd\ and of the lower-mass object 1RXS\,J171502.4$-$333344, respectively, while Section 5 summarizes the results. An appendix yields the properties of the neighbouring HD\,155889.

\section{Observations and data reduction}
The field surrouding \hd\ and 1RXS\,J171502.4$-$333344 was observed by \xmm\ on 2008 Aug. 26 (Rev. 1596, PI Naz\'e), for a total exposure time of 35\,ks. We processed the data with the Science Analysis System (SAS) software, version~8.0; further analysis was performed using the FTOOLS tasks and the XSPEC software v 11.2.0. 

As the brightest star in the field (\hd) is rather optically bright, this observation used the three European photon-imaging cameras (EPICs) in the standard, full-frame mode with a thick filter to reject optical light. The EPIC data were first processed with the pipeline chains (tasks {\sc emproc, epproc}) and then filtered following the recommendations of the SAS team to keep only the best-recorded events (pattern between 0 and 12 + \#XMMEA\_EM filter for the EPIC MOS detectors, pattern between 0 and 4 + flag$=$0 for the EPIC pn detector). The light curve at high energies (pulse invariant channel number$>$10000, or E$\gtrsim$10\,keV, and with pattern$=$0) shows no flare that could have affected our dataset. A detection algorithm (task $edetect\_chain$) was applied to the \xmm\ dataset, with a minimum likelihood of 20. The brightest X-ray source in the field is 1RXS\,J171502.4$-$333344. Comparing the derived list of X-ray sources with the list of hot stars in the field-of-view (FOV) yields only two objects in common: \hd\ and HD\,155889. Both are O-type stars, the only two of the FOV, and none of the 5 B-type stars (HD numbers 155703, 155754, 155888, 155942,  319583) present in the FOV are detected in our exposure. This is not too surprising since B stars, especially the late-type objects, such as those present in the FOV, are known to be fainter X-ray emitters than O stars \citep{ber97,san06}. The count rates of these 3 sources in the 0.4--10\,keV energy band and for different EPIC cameras are provided in Table \ref{tab:ctrate}, together with the hardness ratios $HR_1,HR_2$ defined as (M$-$S)/(M+S) and (H$-$M)/(H+M), respectively, where S, M and H are the EPIC-MOS2 count rates in the 0.4--1.\,keV, 1.--2.\,keV, and 2.--10.\,keV bands, respectively.

   \begin{table*}
      \caption{Hardness ratios and count rates of the studied sources. }
         \label{tab:ctrate}
     \centering
         \begin{tabular}{lccccc}
            \hline\hline
Source & MOS1 & MOS2 & pn & $HR_1$& $HR_2$\\
& cts\,s$^{-1}$ & cts\,s$^{-1}$ & cts\,s$^{-1}$ & &  \\
            \hline
\hd       & 0.070$\pm$0.002 & 0.069$\pm$0.002 & 0.311$\pm$0.003   & --0.63$\pm$0.02 & --0.94$\pm$0.03 \\
HD\,155889&   & 0.032$\pm$0.002 &  & --0.28$\pm$0.05 & --0.44$\pm$0.09 \\
1RXS      & 0.464$\pm$0.004   & 0.484$\pm$0.004   & 1.743$\pm$0.008   & --0.16$\pm$0.01 & --0.56$\pm$0.01 \\
            \hline
         \end{tabular}
   \end{table*}

To analyse the high-energy properties of \hd, a circular source region of 40\arcsec\ radius and a nearby background region of the same shape were used to extract EPIC spectra and lightcurves. Similar regions were used for the bright source 1RXS\,J171502.4$-$333344. For the spectral analyses, the spectra were binned to reach at least 10 counts per bin, and both bad energy bins and noisy bins at E$<$0.3\,keV or $>$10.\,keV were discarded. 

Our observation also consists of data from the reflection grating spectrograph (RGS) for \hd\ and 1RXS\,J171502.4$-$333344 (which is 220\arcsec\ west of \hd). First, the pipeline chain (task {\sc rgsproc}) was run with the coordinates of the objects and processed in a standard way (tasks $rgsspectrum$, $rgsrmfgen$, $rgsfluxer$). No flare in the lightcurve was detected, as was already the case for the EPIC cameras. The RGS spectra were binned to reach at least 5 cts per bin. The data from the second order does not contain enough counts to be useful, so we discarded them from the analysis. A check was made to ensure that one source was not contaminating the other. For the background estimation of one source, the other source was excluded from the background region to avoid background contamination. According to the \xmm\ user manual, the RGS point-spread-function has an $FWHM$ of 0.25\arcmin\ in the cross-dispersion direction, but the sources are separated by 0.8\arcmin\ in this direction, which makes the cross-contamination rather negligible.

\section{\hd, the hottest known Oe star}

\subsection{The Oe category}

The class of Oe stars was officially defined by \citet{con74} as O-stars displaying emission in the Balmer lines without emission in He\,{\sc ii}\,$\lambda$4686 or in the neighbouring N\,{\sc iii} triplet (two signatures of a strong stellar wind). The shape of the Balmer lines in these stars is reminiscent of that in Be objects, and the Oe stars are thus often seen as the continuity of the Be phenomenon to higher temperatures. It is true that most of these stars display high projected rotational velocities $v \sin(i)$ \citep{neg04} but, contrary to Be stars, direct evidence of that disc is still lacking \citep[see e.g.][]{vin09}, although indirect evidence, such as variability similar to that of Be stars, has been observed in the emission lines of several Oe stars \citep{rau07}. It should also be noted that the Oe stars are quite rare: only eight members of this class are known so far and all appear to be of late O-type (O9-B0, \citealt{neg04}), with the sole exceptions of \hd\ (O7.5) and maybe HD\,39680 (O8.5). This is taken as evidence that high temperatures are incompatible with the Be characteristics, a conclusion that agrees with the predictions of \citet{mae00}, who notably point out the influence of strong mass loss and its accompanying loss of angular momentum in the most massive stars.

The X-ray emission of the Oe stars has never been investigated, and their properties in this domain are thus totally unknown, despite the fact that X-rays could help for showing the link (or absence of link) with the more common Be stars. Depending on the exact nature of the Oe phenomenon, different behaviours could be expected. First, if the dense equatorial regions have little impact on the overall wind outflow, the X-ray emission should appear very similar to that of ``normal" O-type stars, i.e. the usual wind-shock model of spherically-symmetric winds would apply. Second, if the dense equatorial regions are optically thick, but the only effect at high-energies concerns absorption, some asymmetries in the line profiles could be observed, depending on the viewing angle. Finally, if the equatorial feature is really dominating the circumstellar environment, hard X-ray emission would be expected. In this context, one may wonder whether hot `$\gamma$\,Cas analogs\footnote{In X-rays, the Be star $\gamma$\,Cas displays variability, a high overluminosity (\loglxlb=$-$5.4), a dominant plasma extremely hot (12\,keV), a strong continuum component, and rather narrow X-ray lines - though not as narrow as for some B stars \citep{smi04}. The disc here, presumably, plays a non-negligible role, as a source of absorption and, possibly, of soft X-ray emission. Similar characteristics were found in a few other emission-line stars \citep{rak06,lop06,lop07}, which prompted the definition of a new subclass of X-ray active Be stars, the so-called `$\gamma$\,Cas analogs'.}' exist or if magnetic confinement plays a role in the putative dense equatorial regions. Indeed, for single O stars (i.e.\ objects without colliding wind emission or emission from accretion by a compact companion), the signature of a magnetic confinement can be distinguished from intrinsic wind shocks. The best known example of such cases is $\theta^1$\,Ori\,C, whose high-energy characteristics (narrow lines, very hot plasma) are well-fitted by magnetohydrodynamical models \citep{gag05}. 

For the first X-ray investigation of Oe stars, we focus on the hottest Oe star, \hd, which had been detected by \ros\ during its all-sky survey \citep{rxs}. First reported as a Be object in 1925 \citep{mer25}, it was later recognised as a true Oe star \citep{hil69,con74} of type O7.5IIIe or V[n]e\footnote{ \citet{neg04} favour the original giant classification of \citet{con74}, but they recognise that the star's luminosity is `on the low side'. This is probably why \cite{wal73} rather used the dwarf classification. In this paper, we follow Negueruela's choice, keeping in mind the above caveat.}. \hd\ indeed displays Balmer lines in emission (with a multiple-peak structure in H$\alpha$, see e.g. Fig. 8 in \citealt{neg04}), but also Fe\,{\sc ii} emission lines \citep{wal80} and double-peaked emissions in C\,{\sc ii}\,$\lambda$4268\AA, Mg\,{\sc ii}\,$\lambda$4481\AA, He\,{\sc i}\,$\lambda\lambda$4713,6678\AA. Its terminal wind velocity and projected rotational velocity were estimated by \citet{how97} to 2460\,\kms\ and 91\,\kms, respectively. Variations in the Balmer lines are common in Oe/Be objects, and \hd\ is no exception (see Fig. A.1 in \citealt{hub08}). However, it is important to stress that, for this star, the \ha\ emission never disappears as the variations only affect the upper part of the line (note that an archival UVES dataset of April 2008, close to the time of the \xmm\ observation, clearly shows the usual, strong \ha\ emission). In addition, no velocity variation was found for this star \citep{gar80}, so binarity is not thought to play a role in its behaviour. Recently, a weak magnetic field has been detected in one spectropolarimetric dataset of \hd: $<Bz>=-115\pm37$\,G \citep{hub07}. Subsequent investigation may have revealed some polarity changes from one day to the next, though these were only marginal detections \citep{hub08} that await confirmation \citep{pet09}. Further observations will be needed to confirm the presence of such a magnetic field in this object.

\subsection{The observed X-ray emission of \hd}

   \begin{table*}
      \caption{Best-fit models and X-ray fluxes at Earth for the two O-type stars. }
         \label{tab:fit}
     \centering
         \begin{tabular}{lcccccccccc}
            \hline\hline
Source & $N_H$ & k$T_1$ & norm$_1$ & k$T_2$ & norm$_2$ & k$T_3$ & norm$_3$ &  $\chi^2_{\nu}$ (dof) & $f_{\rm X}^{\rm abs}$ & $f_{\rm X}^{\rm unabs}$\\
& $10^{22}$\,cm$^{-2}$& keV   & $10^{-4}$cm$^{-5}$ & keV   & $10^{-4}$cm$^{-5}$ & keV   & $10^{-4}$cm$^{-5}$  & & \multicolumn{2}{c}{$10^{-13}$~erg\,cm$^{-2}$\,s$^{-1}$}  \\
            \hline
\vspace*{-0.3cm}&\\
\hd\ (EPIC) & 0.$_{0.}^{0.003}$ & 0.19$_{0.18}^{0.19}$ & 4.51$_{4.34}^{4.63}$ & 0.48$_{0.45}^{0.50}$ & 1.80$_{1.65}^{1.95}$ &  & & 1.54 (410) & 4.78 & 9.16\\
\vspace*{-0.3cm}&\\
\hd\ (EP+RGS) & 0.04$_{0.02}^{0.06}$ & 0.20$_{0.18}^{0.21}$ & 23.5$_{10.1}^{47.0}$ & 0.73$_{0.62}^{0.79}$ & 0.61$_{0.50}^{1.02}$ &  & & 1.36 (869) & 4.61 & 8.69\\
\vspace*{-0.3cm}&\\
\multicolumn{9}{l}{+ abundances compared to solar : N=0.59$_{0.29}^{0.99}$, O=0.21$_{0.15}^{0.36}$, Ne=0.42$_{0.36}^{0.51}$}\\
\vspace*{-0.1cm}&\\
HD\,155889 (EP, 2T) & 0.42$_{0.23}^{0.56}$ & 0.09$_{0.06}^{0.11}$ & 474$_{184}^{1729}$ & 0.61$_{0.50}^{0.71}$ & 2.14$_{1.33}^{3.37}$ &  & & 1.30 (38) & 1.84& 3.14\\
\vspace*{-0.3cm}&\\
HD\,155889 (EP, 3T) & 0.01$_{0.}^{0.27}$ & 0.12$_{0.07}^{0.15}$ & 3.49$_{1.39}^{10.6}$ & 0.63$_{0.46}^{0.72}$ & 0.60$_{0.46}^{4.17}$ & 3.02$_{1.93}^{9.12}$ & 0.65$_{0.29}^{0.91}$& 1.20 (36) & 2.21 & 3.49\\
\vspace*{-0.3cm}&\\
            \hline
         \end{tabular}
   \end{table*}

In the \xmm\ data, the X-ray spectrum of \hd\ immediately appears soft, with basically no counts detected above 2\,keV (Fig.\ref{rgs}). It is thermal in nature, as proven by the appearance of the RGS spectrum (see below), and the EPIC data were thus fitted by a sum of absorbed optically-thin thermal components. Two temperatures were needed to obtain a good fit, and the models thus take the form $wabs(ISM)\times wabs\times(apec+apec)$, with $N_H^{\rm ISM}$ fixed to 1.23$\times10^{21}$~cm$^{-2}$ \citep[ in agreement with the value derived from the $B-V$ colour excess]{dip94}. Table \ref{tab:fit} lists the derived spectral parameters in that case: the indices and exponents noted for each parameter correspond to the lower and upper limits of the 90\% confidence interval (derived from the $error$ command under {\sc xspec}), respectively; the normalisation factors are defined as $10^{-14}\int n_e n_{\rm H} dV/4\pi D^2$, where $D$, $n_e$, and $n_{\rm H}$ are respectively the distance to the source, the electron, and proton density of the emitting plasma; the abundances were kept solar, and the fluxes are given in the 0.5--10\,keV band. Using the bolometric corrections of \citet{mar06} for an O7.5III star, a distance to Sco\,OB4 of 1.1\,kpc \citep{pis07}, and a V magnitude of 5.61, the bolometric luminosity was evaluated to $7.4\times10^{38}$\ergs\ and the \lxlb\ ratio is therefore $-6.75$. This value is quite normal for an O-type star \citep{naz09}, and there is thus no X-ray overluminosity for \hd.

\begin{figure*}
\begin{center}
\includegraphics[width=9.5cm]{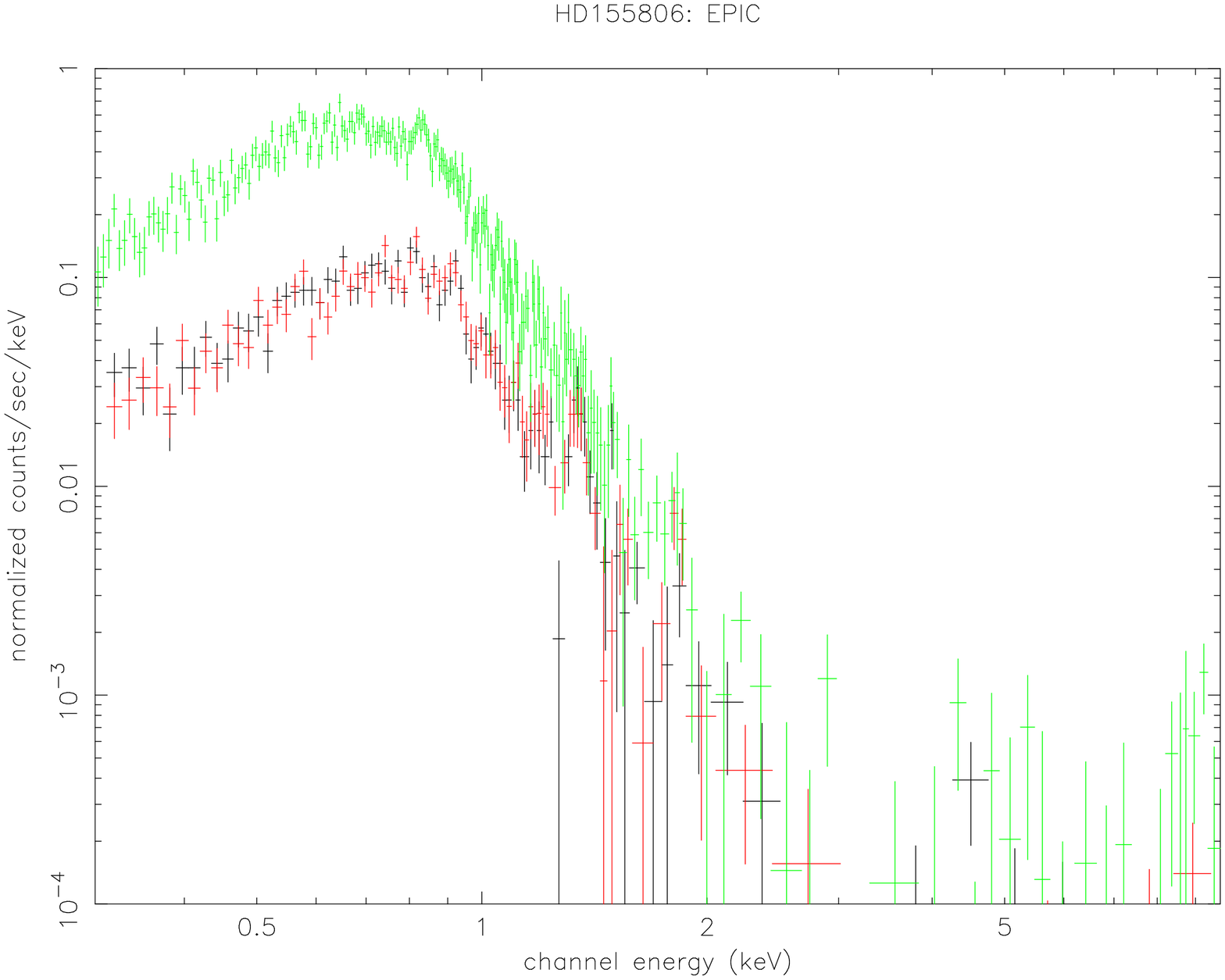}
\includegraphics[width=7.5cm,bb=45 200 575 700,clip]{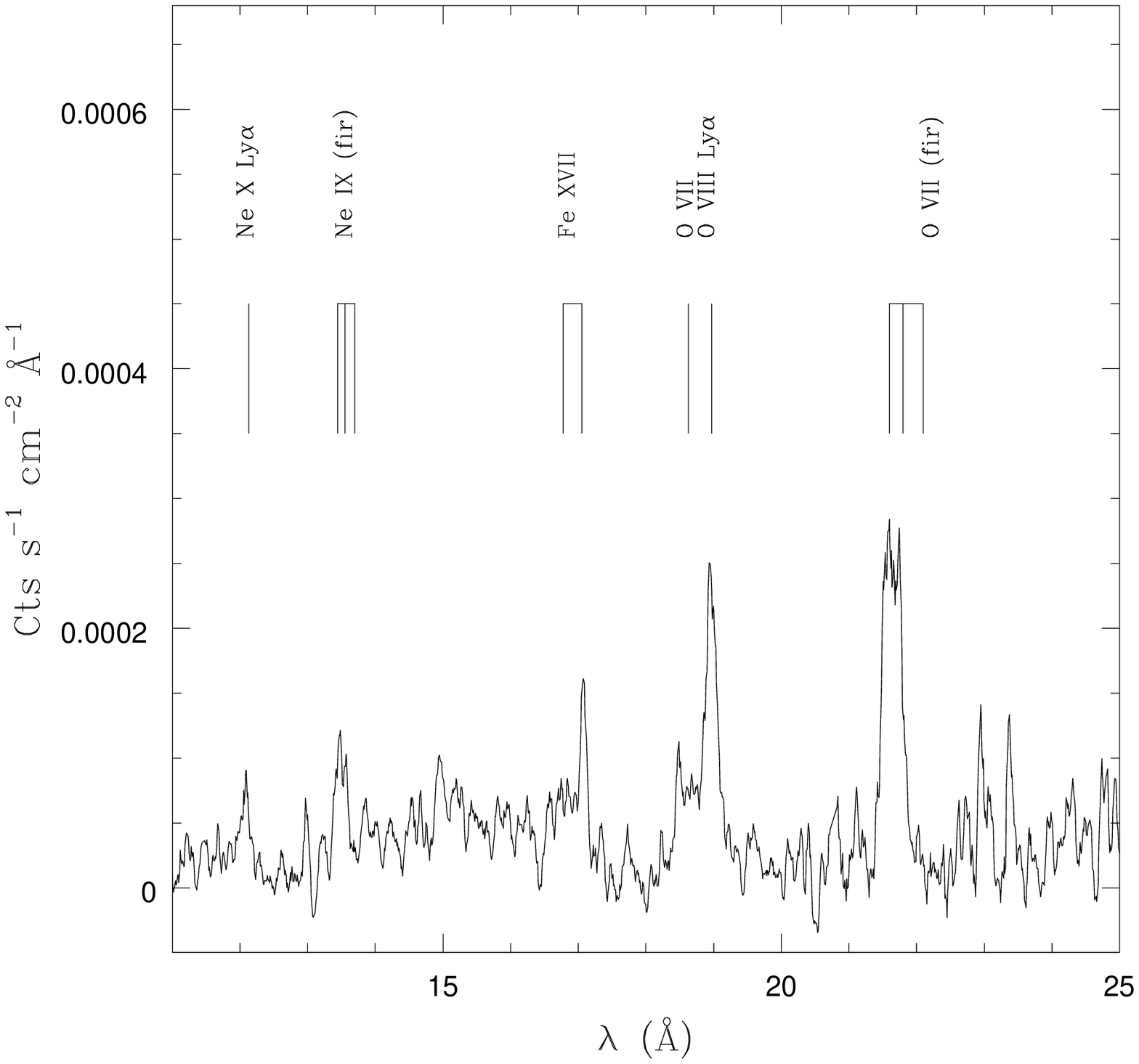}
\caption{\label{rgs} Left: EPIC spectra of \hd\ (pn data shown as the upper green curve, MOS data as the lower red and black curves). Right: Fluxed RGS spectrum of \hd\ (from task {\sc rgsfluxer}, RGS1+2, orders 1+2) with the main lines identified. The data were smoothed using a sliding box, to highlight the X-ray lines.} 
\end{center}
\end{figure*}

Because of the relative faintness of \hd, the high-resolution X-ray spectrum is quite noisy and only a few limited conclusions can be drawn from it. Some X-ray lines are clearly detected (Fig. \ref{rgs}): Lyman-$\alpha$ lines of hydrogen-like neon and oxygen (Ne\,{\sc x}\,$\lambda$12.13\AA, O\,{\sc viii}\,$\lambda$18.97\AA), as well as $fir$ triplets of He-like ions of the same elements (Ne\,{\sc ix}\,$\lambda$13.5\AA, O\,{\sc vii}\,$\lambda$21.8\AA). For these triplets, the $f$ line clearly has a much lower amplitude than the $r$ and $i$ lines, as expected for such a hot star emitting a wealth of UV photons \citep[see ][for a discussion of this effect]{gud09}. The noise prevents us from deriving strong constraints from them, but only yields $f/i<0.1$, corresponding to an X-ray source situated below $\sim$10\,R$_*$ from the stellar photosphere \citep{raa08}, for the brightest triplet, that of O\,{\sc vii}. A simultaneous fit of the EPIC and RGS data provides hints of reduced abundances in N, O, and Ne (Table \ref{tab:fit}), but these values should be independently confirmed because of the large uncertainties introduced by the noise in the RGS data.  

The strongest lines, O\,{\sc viii}\,Ly$\alpha$ and the triplets of O\,{\sc vii}, were the sole ones to provide enough counts to be fitted using Gaussians with a fixed central energy (SPEX v2.0: 0.6533\,keV for O\,{\sc viii}\,Ly$\alpha$ and 0.56101, 0.56874, 0.57395\,keV for the O\,{\sc vii} triplet). The resulting line widths were estimated to $FWHM=2500_{1700}^{3800}$\,\kms\ for O\,{\sc viii}\,Ly$\alpha$ and $FWHM=3200_{2100}^{4200}$\,\kms\ for the O\,{\sc vii} triplet, i.e. values comparable to the terminal velocity (2460\,\kms, \citealt{how97}). A slightly poorer fit was found for O\,{\sc viii}\,Ly$\alpha$ in the case of a width of 1400\,\kms, with a lower limit of the confidence interval of about 1000\,\kms. Small widths, $<$400\,\kms\ like those of $\gamma$\,Cas \citep{smi04}, $\theta^1$\,Ori\,C \citep{gag05}, or $\theta$\,Car \citep{naz08}, are clearly excluded. 

\begin{figure}
\begin{center}
\includegraphics[width=8cm]{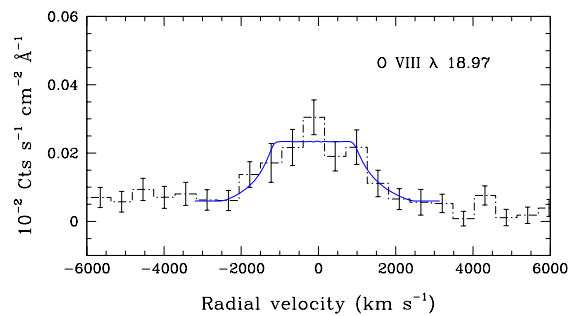}
\caption{\label{line} The best-fit exospheric line-profile model (with the X-rays emitted above $R_0 = 1.85$\,$R_*$ and a wind opacity of $\tau_{\lambda,*}$ = 0.0) overplotted on the RGS data of the O\,{\sc viii}\,Ly$\alpha$ line (binned to get 1000 bins for the entire wavelength range).} 
\end{center}
\end{figure}

As these properties are typical of ``normal" O-type stars, we further attempted to fit the O\,{\sc viii} Ly$\alpha$ line with an exospheric line profile model following the formalism of \citet{owo01}. In this model, the X-ray emission originates in material distributed throughout a homogeneous spherical wind, above a radius $r \geq R_0 > R_*$, and the hot plasma follows the same velocity law as the cool wind. Doppler broadening due to macroscopic motion therefore provides the main source of line broadening. The line emissivity is assumed to scale as $\epsilon \propto \rho^2$, which is equivalent to the assumption of a constant filling factor of the X-ray plasma throughout the wind. The free parameters of this model are thus $R_0$ and $\tau_{\lambda,*} = \frac{\kappa_{\lambda}\,\dot{M}}{4\,\pi\,v_{\infty}\,R_*}$, the characteristic optical depth at wavelength $\lambda$. For the terminal wind velocity, we adopted $v_{\infty} = 2460$\,km\,s$^{-1}$ as derived by \citet{how97}. In our fits, we accounted for a (pseudo)-continuum beneath the oxygen line with a level of $6 \times 10^{-5}$\,cts\,s$^{-1}$\,cm$^{-2}$\,\AA$^{-1}$. The best fit to the line profile with default binning (resulting from the task {\sc rgsfluxer}: 3400 bins between 4 and 38.2\AA) is obtained for $R_0 \simeq 2.05$\,$R_*$ and $\tau_{\lambda,*} = 0$ corresponding to a flat-topped profile. The 90\% confidence intervals on these parameters are approximately $[1.5,3.0]$ and $[0.0,1.0]$ for $R_0$ and $\tau_{\lambda}$, respectively. The fits to a higher signal-to-noise fluxed spectrum composed of 1000 bins nicely confirm this picture, though the best-fit value of $R_0$ is now $\simeq 1.85$\,$R_*$ (Fig. \ref{line}). These values of the opacity and onset radius are not exceptional amongst hot stars. \citet{leu06} found $R_0=1.25-1.67$\,R$_*$ and $\tau=0-1$ for a sample of closeby, X-ray bright O-type stars. Though their X-ray spectra had a much better signal-to-noise than in our study, we may note that the parameters of $\zeta$\,Ori resemble most those of \hd. 

A final comment can be made on the (a)symmetry of the X-ray lines. This property is important in the debated question of the origin of the X-ray emission from massive stars, with notably dense winds producing blueward-skewed lines and porous winds more symmetric lines \citep[see][ for a review]{gud09}. For \hd, the sole line that is both isolated and strong enough to perform such an analysis is the O\,{\sc viii}\,Ly$\alpha$ line. First, the Gaussian fitting within Xspec shows no significant trend in the residual that would indicate an asymmetry  \citep[for a contrasting case, see][]{coh06}. It is also obvious from the above fitting that the best exospheric model is symmetric ($\tau$=0). Finally, the first three moments were calculated for the RGS spectrum derived from the task {\sc rgsfluxer} on the interval $[-v_{\infty}..v_{\infty}]$ \citep[as was also done on Chandra data by e.g. in][ but we do normalise the 3rd-order moment]{coh06}. The first moment indicates no line shift: $\mu_1=-0.03$ to $-0.09$ with a 1-$\sigma$ error of 0.045 and the x-axis defined as $\left( \frac{\lambda}{\lambda_0}-1 \right) \frac{c}{v_{\infty}}$. The second- and third-order moments, calculated with respect to the rest wavelengths of the Ly$\alpha$ components (i.e. 18.977 and 18.971\AA) and for the different binnings (default 3400 binning, custom 1000 binning), yield $\mu_2=0.21\pm0.05$ (a result in agreement with the FWHM calculated above) and $\mu_3=-0.02$ to $-0.06\pm0.05$, i.e. not significantly different from zero. Equivalently, the derived skewness is $\mu_3/\mu_2^{3/2}=-0.2$ to $-0.6\pm0.5$. Within the limitations of our data, the lines of \hd\ should thus be considered symmetric. However, we note that (1) because of the noise in our data, only large asymmetries would have been detected, if present; and (2) even for bright sources with much better RGS spectra, conflicting results are sometimes found \citep[see e.g. the case of $\zeta$\,Ori in ][vs. \citealt{coh06,pol07}]{wal01,raa08}.

Over recent years, it has been shown that the assumption of a homogeneous wind might not be entirely appropriate for treating the absorption of the X-ray emission by the stellar wind of O-type stars and that clumping or porosity \citep{osk04,owo06} likely affect the wind opacity. This is probably why our best-fit optical depth is consistent with 0.0. However, given the limited quality of our data, we preferred to fit the profile with a simple homogeneous wind model. 

The \xmm\ lightcurve was investigated with $\chi^2$ and $pov$ tests \citep{san04}. No significant variations were detected. For investigating long-term variations, we can only compare our \xmm\ observation with the sole previous \ros\ detection. Within the RASS faint source catalogue \citep{rxs}, \hd\ is recorded as 1RXS J171518.9$-$333250. This source presented a PSPC count rate of (4.40$\pm$1.45)$\times$10$^{-2}$\,cts\,s$^{-1}$.  The best-fit model from Table \ref{tab:fit} predicts a \ros-PSPC count rate of 4.7$\times$10$^{-2}$\,cts\,s$^{-1}$, which agrees with the previous detection. To the best of our knowledge, \hd\ therefore appears non-variable at X-ray energies.

A soft X-ray spectrum, broad X-ray lines, the absence of X-ray overluminosity, and lack of variations: all the evidence agrees with the `classical' wind-shock model as the origin of the high-energy emission. \hd\ resembles `normal' O-stars, such as HD\,155889 (see Appendix), and is certainly not a hot $\gamma$\,Cas analog, though one Oe star apparently belongs to this class HD\,119682 (O9.7e, \citealt{rak06}): it is still unclear why some Oe/Be do present similarities with $\gamma$\,Cas, while others do not. A dedicated multiwavelength survey of a large sample of such objects would help to better constrain the physical parameters responsible for the various Oe/Be phenomena.

In addition, our observations do not indicate a strong contribution from magnetic phenomena, such as magnetically-confined winds (exemplified by $\theta^1$\,Ori\,C). On the basis of the single magnetic detection of \citet{hub07}, the estimated magnetic confinement parameter \citep[see][ for a full definition]{udd02} for this star is about 0.4, i.e. quite close to 1, the limit where magnetic phenomena dominate. However, changes in polarity were reported by \citet{hub07}, and the single observed value, rather uncertain, thus might not reflect the actual strength of the overall stellar magnetic field of \hd. To explain the X-ray characteristics and pinpoint the magnetic field properties, a new, sensitive spectropolarimetry campaign covering the likely rotation period of the star, i.e. spanning several days, should be undertaken. 

In conclusion, in the case of \hd, the putative dense equatorial region apparently plays a minor role in the X-ray emission. 

\section{1RXS\,J171502.4$-$333344}

The brightest source in the field is neither \hd\ nor HD\,155889 but a nearby source detected previously by \ros: 1RXS\,J171502.4$-$333344 (Table \ref{tab:ctrate}). As it has never been studied in detail, it is valuable to present here the results of a high-quality observation of this source. Its coordinates (Eq. 2000.0) are refined to 17h15m2.3s $-$33$^{\circ}$33\arcmin43.2\arcsec. The formal error on the coordinates is 0.03\arcsec, but we note that the position of \hd\ in the X-ray data is offset by 1.3\arcsec\ with respect to the GSC2.3 coordinates of the latter star. The closest counterpart is the star catalogued as USNO-B1.0 0564--0551957 or 2MASS 17150219--3333398, with magnitudes $R\sim$10.3, $I$=8.6, $J$=7.92, $H$=7.28, $K$=7.07. The \ros\ source was indeed considered associated with this 2MASS objet \citep{ria06}, but we note that it is situated at 3.7-3.8\arcsec\ from the XMM source, which is a large positional error for an X-ray source close to the centre of the FOV. However, the 2MASS star is situated at a distance of only 31pc, so it displays a relatively high tangential velocity of 25\,\kms\ \citep{ria06}, which could (at least partially) account for the separation between the 2MASS coordinates of this object and the position of the XMM source. 

   \begin{table*}
      \caption{Same as Table \ref{tab:fit} for 1RXS\,J171502.4$-$333344. }
         \label{tab:fit1RXS}
     \centering
         \begin{tabular}{lcccccccccc}
            \hline\hline
Source & $N_H$ & k$T_1$ & norm$_1$ & k$T_2$ & norm$_2$ & k$T_3$ & norm$_3$ &  $\chi^2_{\nu}$ (dof) & $f_{\rm X}^{\rm abs}$ & $f_{\rm X}^{\rm unabs}$\\
& $10^{22}$\,cm$^{-2}$& keV   & $10^{-4}$cm$^{-5}$ & keV   & $10^{-4}$cm$^{-5}$ & keV   & $10^{-4}$cm$^{-5}$  & & \multicolumn{2}{c}{$10^{-13}$~erg\,cm$^{-2}$\,s$^{-1}$}  \\
            \hline
1RXS (EP, quiet, Z=0.24$_{0.22}^{0.25}$) & 0.$_{0.}^{0.002}$ & 0.30$_{0.28}^{0.31}$ & 11.5$_{10.5}^{12.8}$ & 0.78$_{0.76}^{0.81}$ & 11.9$_{11.0}^{13.3}$ & 1.48$_{1.41}^{1.56}$ & 9.53$_{8.70}^{10.2}$& 1.21 (768) & 21.9  & 21.9\\
\vspace*{-0.3cm}&\\
1RXS (EP, flare, Z=0.40$_{0.32}^{0.45}$) & 0.$_{0.}^{0.003}$ & 0.33$_{0.32}^{0.37}$ & 16.5$_{14.8}^{19.4}$ & 0.98$_{0.94}^{1.02}$ & 21.9$_{18.9}^{28.3}$ & 3.00$_{2.83}^{3.23}$ & 53.0$_{49.9}^{55.6}$& 0.94 (850) & 98.7  & 98.7 \\
\vspace*{-0.3cm}&\\
1RXS (EP+RGS, Z=0.24$_{0.20}^{0.26}$)& 0.$_{0.}^{0.003}$ & 0.33$_{0.32}^{0.33}$ & 15.0$_{14.2}^{18.5}$ & 0.95$_{0.92}^{0.96}$ & 16.2$_{14.7}^{20.6}$ & 2.54$_{2.39}^{3.09}$ & 12.4$_{9.80}^{13.5}$& 1.17 (1816) &  32.7 & 32.7\\
\vspace*{-0.3cm}&\\
1RXS (EP, quiet) & 0.$_{0.}^{0.004}$ & 0.25$_{0.24}^{0.26}$ & 8.47$_{7.01}^{10.2}$ & 0.63$_{0.61}^{0.64}$ & 13.0$_{11.3}^{15.1}$ & 1.51$_{1.44}^{1.58}$ & 10.56$_{9.66}^{11.4}$& 1.09 (765) & 22.0  & 22.0\\
\vspace*{-0.3cm}&\\
\multicolumn{9}{l}{+ abundances compared to solar : CNOS=0.23$_{0.19}^{0.26}$, MgAlSi=0.24$_{0.19}^{0.29}$, CaFeNi=0.19$_{0.17}^{0.21}$, NeAr=0.82$_{0.72}^{0.91}$}\\
\vspace*{-0.3cm}&\\
1RXS (EP, flare) & 0.$_{0.}^{0.007}$ & 0.31$_{0.29}^{0.33}$ & 20.8$_{17.1}^{29.2}$ & 0.99$_{0.94}^{1.04}$ & 16.4$_{13.0}^{21.7}$ & 2.87$_{2.61}^{3.05}$ & 57.3$_{54.2}^{60.4}$& 0.92 (847) & 98.7  & 98.7 \\
\vspace*{-0.3cm}&\\
\multicolumn{9}{l}{+ abundances compared to solar : CNOS=0.28$_{0.20}^{0.37}$, MgAlSi=0.29$_{0.14}^{0.44}$, CaFeNi=0.44$_{0.36}^{0.50}$, NeAr=0.74$_{0.49}^{1.01}$}\\
\vspace*{-0.3cm}&\\
1RXS (EP+RGS)& 0.$_{0.}^{0.001}$ & 0.26$_{0.25}^{0.27}$ & 7.30$_{6.28}^{8.24}$ & 0.64$_{0.63}^{0.65}$ & 13.4$_{12.1}^{15.1}$ & 1.93$_{1.88}^{1.99}$ & 18.4$_{17.7}^{19.1}$& 1.08 (1813) &  32.5 & 32.5\\
\vspace*{-0.3cm}&\\
\multicolumn{9}{l}{+ abundances compared to solar : CNOS=0.29$_{0.26}^{0.31}$, MgAlSi=0.25$_{0.20}^{0.29}$, CaFeNi=0.22$_{0.20}^{0.24}$, NeAr=1.05$_{0.97}^{1.14}$}\\
\vspace*{-0.3cm}&\\
            \hline
         \end{tabular}
   \end{table*}

\begin{figure}
\begin{center}
\includegraphics[width=8cm]{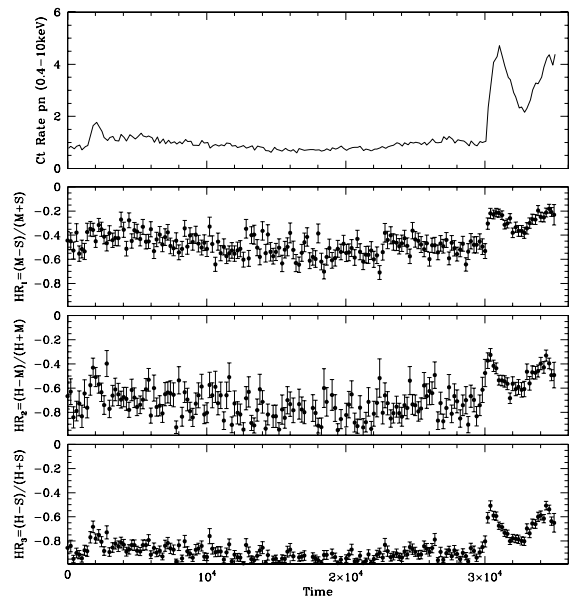}
\caption{\label{1RXSlc} The evolution of the EPIC-pn count rate (top) and hardness ratios for 1RXS\,J171502.4$-$333344, shown with a time bin of 200s.} 
\end{center}
\end{figure}

\begin{figure*}
\begin{center}
\includegraphics[width=9.5cm]{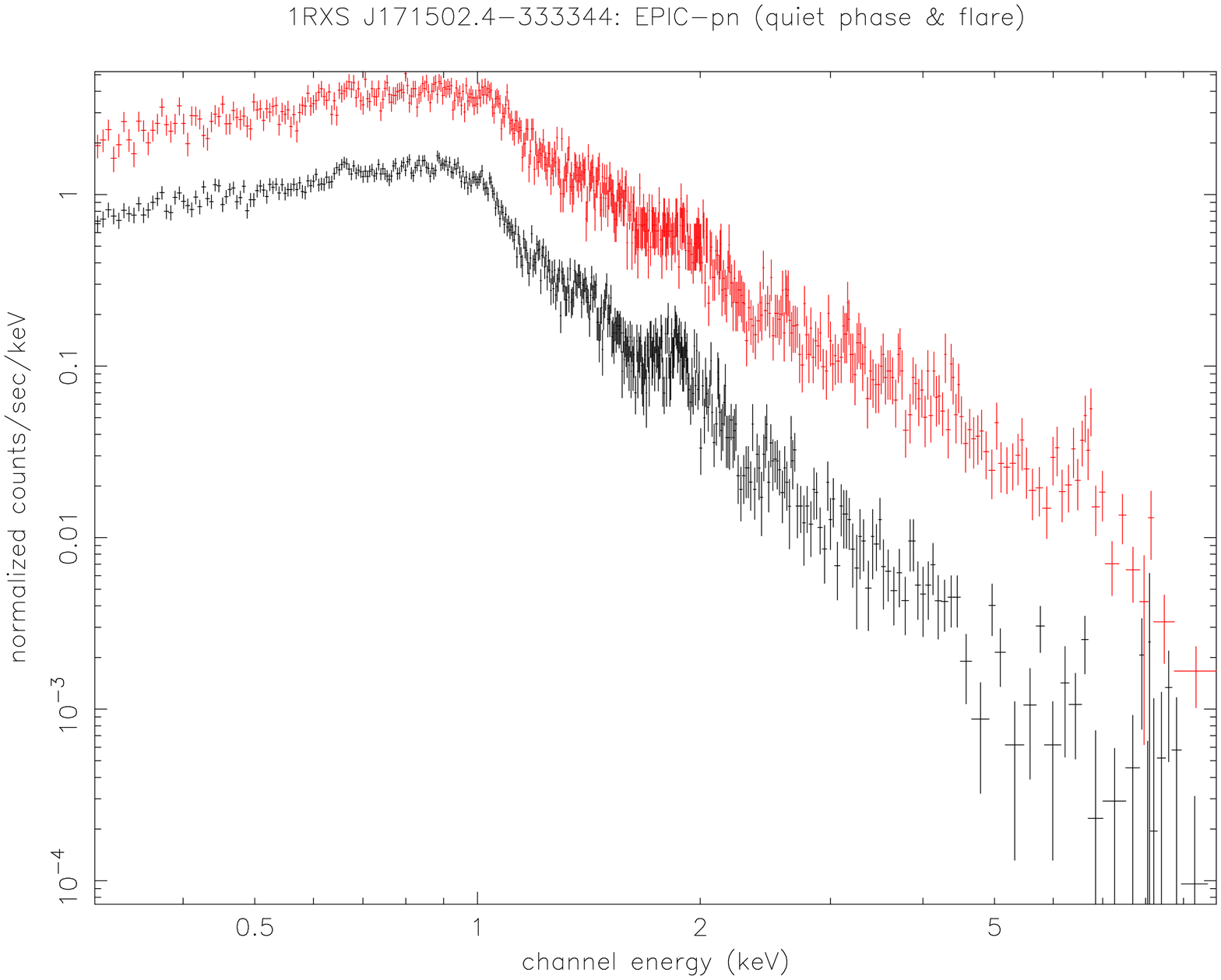}
\includegraphics[width=8cm,bb=45 200 590 700,clip]{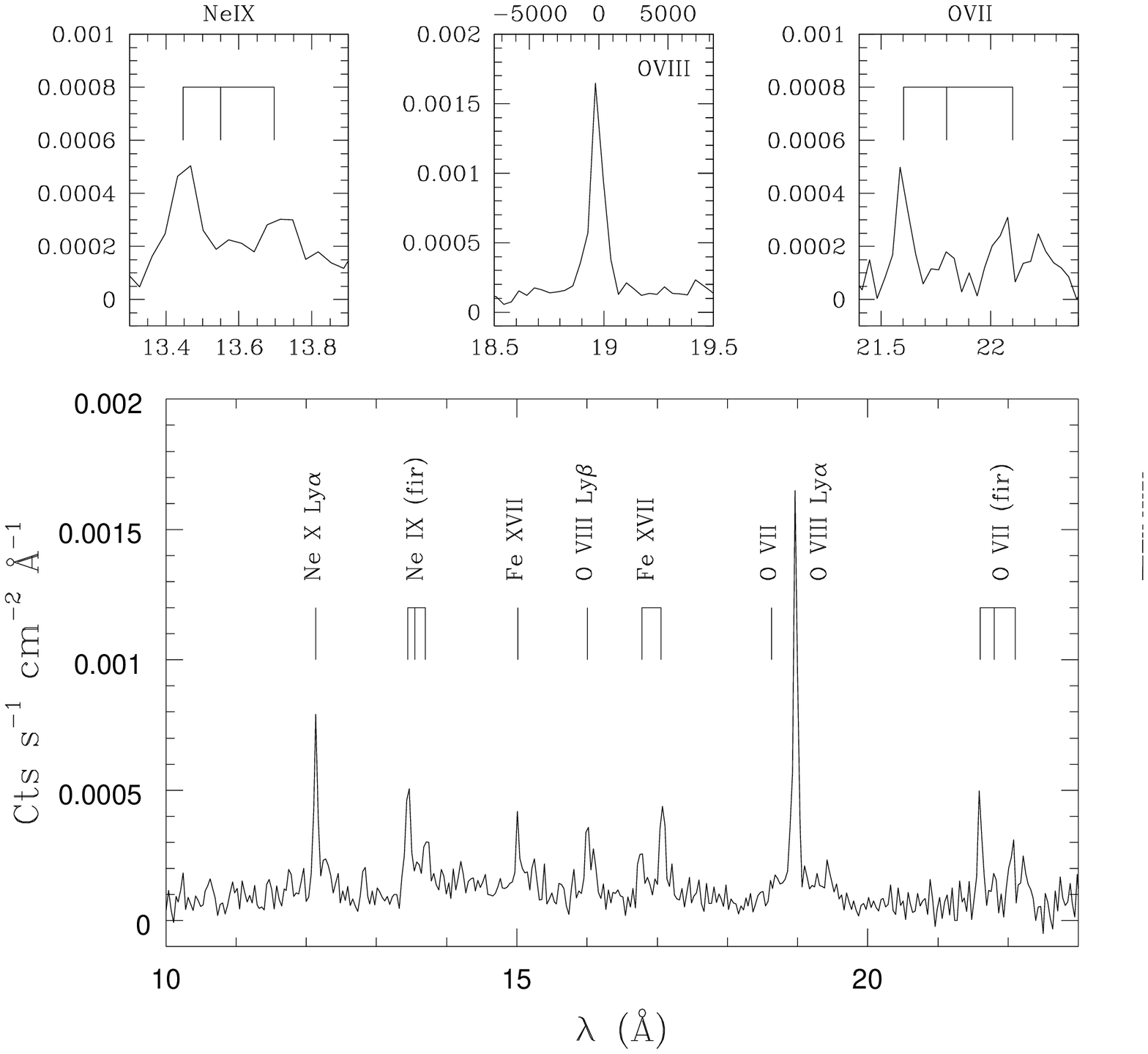}
\caption{\label{1RXS} Left: The EPIC-MOS2 spectra of 1RXS\,J171502.4$-$333344 (upper curve: during the flare, lower curve: in quiescence). Right: fluxed RGS spectrum.} 
\end{center}
\end{figure*}

The lightcurve of 1RXS\,J171502.4$-$333344 sharply contrasts with those of the O-type stars, and instead resembles the lightcurve of AT\,Mic \citep{raa03}. It displays a large flare towards the end of the 35ks observation and a modulation of much smaller amplitude during the whole exposure (plus a very small flaring event near the beginning of the exposure, see Fig. \ref{1RXS}). Though the source is then bright, the EPIC-pn data are not affected by pile-up, and the contamination of the EPIC-MOS data is very limited, as was checked by a run of the task $epatplot$. Unfortunately, the decay phase has not been observed by \xmm\ but it is interesting to note that the rise of the flare presents a much steeper slope than the beginning of the decaying phase. In addition, the hardness ratios are variable, with a small, but definite, hardening during the flare. Finally, as for Ross 154 \citep{war08}, the hard X-ray lightcurve peaks before that in the soft X-ray band. This is reminiscent  of the Neupert effect, which is explained by the radio, optical, and hard X-ray emissions tracing the rate of energy deposition, while the soft X-rays scales with the cumulative thermal energy in the beginning of the flare. This effect has been observed in the Sun and a few other coronal sources \citep{war08}, 1RXS\,J171502.4$-$333344 thus being a new, rare example of such cases. The optical monitor being closed during the exposure, no optical/UV lightcurve is available.

Since the source is close to the optical axis, its RGS spectrum could be extracted, and it clearly reveals the presence of lines, without a strong continuum (Fig. \ref{1RXS}): the X-ray emission thus appears thermal in nature. The fitting procedure was similar to that of the O-type stars, but a faint high-energy tail required an additional, third thermal component to get a good fit to the data. The EPIC spectra before (i.e. the first 30ks of the observation) and during the flare were at first analysed separately (Table \ref{tab:fit1RXS}). As was indicated by the lightcurve, the spectra are indeed brighter (a factor of 5 increase in flux) and harder (the two hardest components now dominate, the soft component remaining relatively unchanged) during the flare. 

\begin{figure}
\begin{center}
\includegraphics[width=8cm]{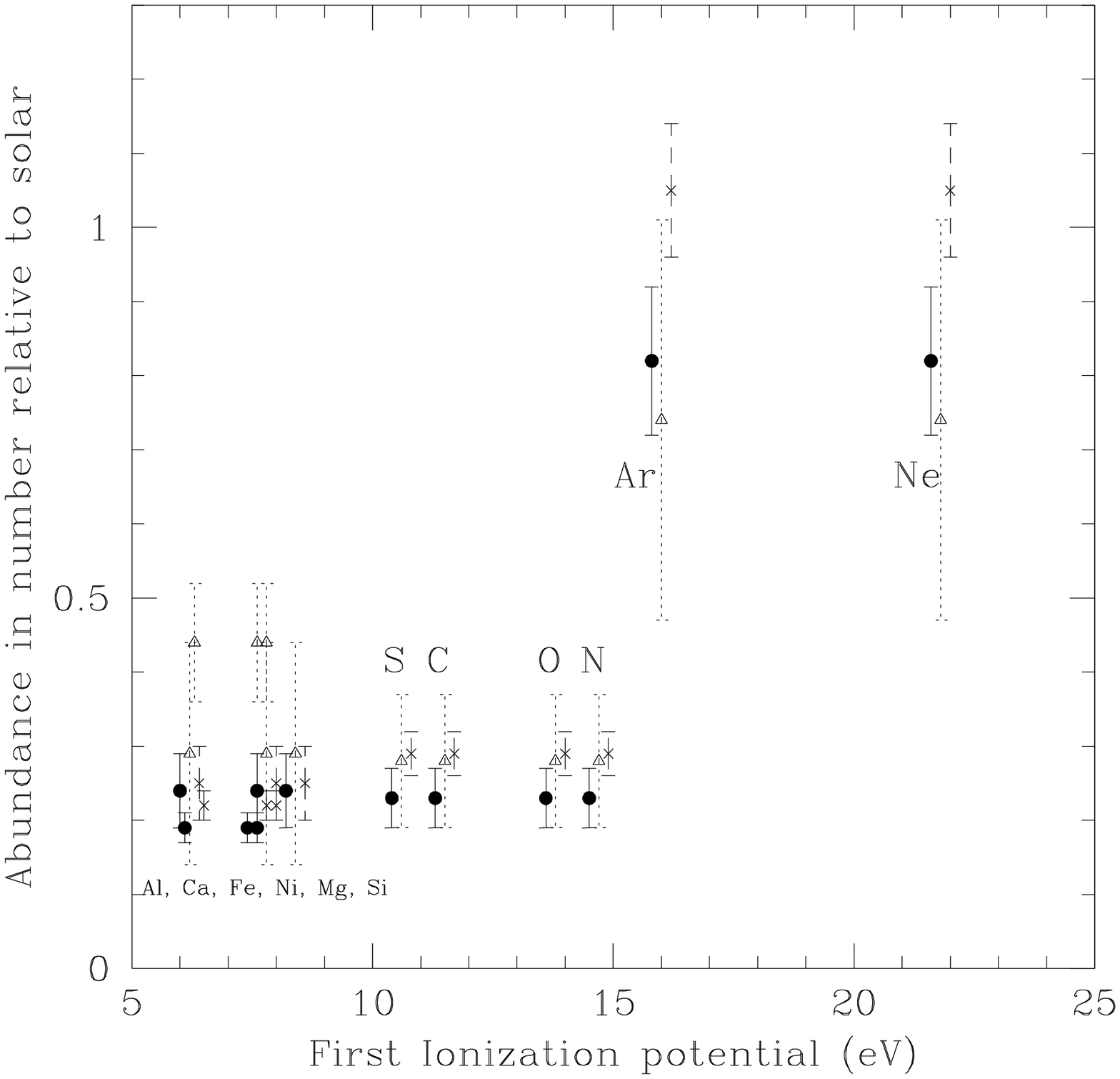}
\caption{\label{abund} The abundances of the elements, ordered by increasing FIP, derived from fits to the X-ray spectrum of 1RXS\,J171502.4$-$333344 (see Table \ref{tab:fit1RXS}). The values derived from EPIC+RGS fits are shown by filled circles, whereas the values associated with the quiescent and flaring phases are shown by open triangles and crosses, respectively (for clarity, they have been shifted right by 0.2 and 0.4 eV, respectively).} 
\end{center}
\end{figure}

One remarkable feature of the fits is the low overall metallicity needed to reproduce the spectrum at all times, with a larger abundance needed during the flare (Table \ref{tab:fit1RXS}). We thus decided to investigate this property further by fitting abundances of groups of elements, i.e. elements with similar first ionization potential (FIP). As in \citet{war08}, we grouped together CNOS, MgAlSi, CaFeNi, and NeAr. The first three groups present similar abundances, but they are definitely lower than the abundances of the fourth group, NeAr, the two elements which have the highest FIPs. Also, the abundances do not vary significantly between the quiescence and the flaring state (Fig. \ref{abund}), except maybe for a slight increase in the CaFeNi group during the flare (though its value is still marginally consistent with the abundance of CNOS and MgAlSi). A simultaneous fitting of the whole EPIC and RGS (order 1) datasets was also undertaken, yielding average properties (Table \ref{tab:fit1RXS} and Fig. \ref{abund}).

As the change in the overall spectral shape is only detectable in the EPIC data at the highest energies ($>$2\,keV), the RGS data remain largely unaffected by the flare, and the whole exposure was thus analysed, to get the best signal-to-noise ratio (Fig. \ref{1RXS}). The $fir$ triplets appear quite different from those of the O-type stars, with a quite strong $f$ line: there is thus no bright source of UV photons in the vicinity of the X-ray emitting regions, nor are they of high density \citep{gud09}. Unfortunately, the noise prevents us from performing a quantitative analysis of these triplets. However, as for \hd, the strongest line, O\,{\sc viii}\,Ly$\alpha$, was fitted by a Gaussian with a fixed central energy (0.6533\,keV). The resulting line width was estimated to $FWHM=180_{60}^{370}$\,\kms.

Given the Galactic coordinates of our field ($l_{II} = 352.59^{\circ}$, $b_{II} = +2.87^{\circ}$) and the fact that the X-ray spectrum of source 1RXS\,J171502.4--333344 shows no evidence of a substantial absorption, we can exclude the possibility that this source could be associated with a background object. On the other hand, the 2MASS colours of the nearest counterpart candidate of this source are actually consistent with the near-IR source being an early M-type star (it was in fact classified as type M0, \citealt{ria06}), whose red spectrum shows signs of mild chromospheric activity ($EW[{\rm H}_{\alpha}]=3$\AA, \citealt{ria06}). 

Several points indeed hint at this X-ray source being a coronally-active dMe object. First, the flare, since magnetically active M dwarves frequently present flares in their X-ray emission \citep[e.g.][]{raa03,rob05,war08}. The hardening during the flare (associated with a higher emission measure of the hottest components) and the shifted maximum of the hard X-ray lightcurve are also indicative of a coronal source \citep{war08}. Second, the luminosity. Adopting the above estimate of the distance, we evaluate a quiescence luminosity of $2.5 \times 10^{29}$\,erg\,s$^{-1}$. This value increases to $1.1 \times 10^{30}$\,erg\,s$^{-1}$ during the flare. The quiescence luminosity is very similar to the typical $L_{\rm X}$ of other dMe stars \citep{mit05,rob05}, and is lower than for PMS T\,Tauri stars or RS\,CVn interacting binaries. Moreover, the $\log(L_{\rm X}/L_{\rm BOL})$ ratio amounts to $-3.3$ in quiescence and $-2.6$ during the flare, which is the typical ratio observed at the saturation limit for coronally-active sources \citep{ria06}. 

Finally, the observed  peculiar abundance pattern is also found in other dMe stars \citep[and references therein]{war08}. The overall subsolar abundance is typical of nearby M stars, though a detailed comparison with the particular photospheric abundances of 1RXS\,J171502.4--333344 must wait for new, high-resolution spectroscopy in the optical range. The increase in the coronal metallicity during the flare has also been observed in some cases, e.g. Ross 145, and it might well reflect the arrival of fresh chromospheric or photospheric material \citep{war08}. The abundance distribution appears flat from Si to N (i.e. the abundances ratios are similar to the solar mixture), but shows a pronounced enhancement for the high-FIP element Ne, with few changes during the flare. Again, this agrees well with observations of other dMe sources \citep{rob05}.

\section{Conclusions}
Our \xmm\ observations of \hd\ and 1RXS\,J171502.4$-$333344 constitute the first X-ray investigations of these objects. 

For \hd, the hottest Oe star, the X-ray emission was found to be quite soft, even slightly softer than for the neighbouring `normal' O-type star HD\,155889: the best-fit thermal components have temperatures of 0.2 and 0.6\,keV, and only one tenth of the intrinsic flux is emitted in the 1.-2.5\,keV energy band. Moreover, the high-resolution spectrum, though noisy, reveals broad, symmetric X-ray lines ($FWHM>1000$\,\kms), and variations were found neither on short (35ks) nor on long (10yrs) timescales. 

The characteristics of the X-ray emission of \hd\ are fully compatible with predictions of the wind-shock model. No signature of magnetic phenomena, such as magnetically-confined winds, was detected. If magnetic fields play a role in Oe stars, it must be a small one, with a reduced impact on the stellar wind and associated X-ray emission. \hd\ does not seem to be a hot $\gamma$\,Cas analog either. Quite surprisingly, its X-ray emission thus appears independent of the dense equatorial regions thought to be linked to the Oe phenomenon.


The characteristics of the source 1RXS\,J171502.4$-$333344, discovered by \ros, are now better constrained: its position was refined, a flare was detected, and its spectra (in quiescence and during the flare) were analysed in detail. High-resolution spectra reveal the presence of narrow X-ray lines and of a strong $f$ line in the triplet of the He-like ion O\,{\sc vii}, whereas lower-resolution data unveil the hardening of this source during the flare. This coronal source, associated to a dMe star, also displays an inverse FIP effect. The observed characteristics of 1RXS\,J171502.4$-$333344 are quite similar to those reported for other dMe stars.

\begin{acknowledgements}
YN and GR acknowledge support from the Fonds National de la Recherche Scientifique (Belgium), the PRODEX XMM and Integral contracts, and the `Action de Recherche Concert\'ee' (CFWB-Acad\'emie Wallonie Europe). AuD acknowledges NASA Grant NNX08AY04G. ADS and CDS were used for preparing this document.
\end{acknowledgements}

\appendix
\section{The neighbouring star HD\,155889 as an example of a ``normal" O-star}
HD\,155889 (O9IV) is a poorly studied, hot star situated some 13\arcmin\ southeast of \hd. \citet{gar80} did not find any velocity variation and concluded that the star is likely to be single. However, an optical companion exists at 0.19\arcsec, and was demonstrated to be a true, physical companion in an elliptical, $>$500\,yrs orbit \citep{mas98,tur08}. The nature of this companion is unknown, but it is about 1\,mag fainter than HD\,155889 in the I-band. The observed colours of HD\,155889 (B=6.51, V=6.57) reveal an extinction similar to that of \hd: the same interstellar absorbing column was thus used for this star.

\begin{figure}
\begin{center}
\includegraphics[width=6cm,angle=270]{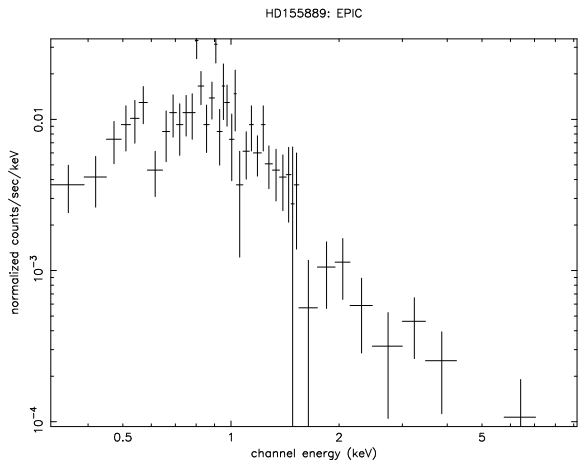}
\caption{\label{hd155889} EPIC-MOS2 spectrum of HD\,155889. } 
\end{center}
\end{figure}

The extraction was more difficult for this star than for \hd\ since the X-ray counterpart appears at the edge of the FOV. Only half of it can be seen with EPIC-MOS1, therefore EPIC spectra and lightcurves were only extracted in EPIC-MOS2 (pn data were unusable). The regions used were a circular source region with a radius reduced to 30\arcsec\ for MOS2 (because of crowding in the area) and a nearby background region of 30\arcsec\ radius. 

Overall, the EPIC-MOS2 spectrum of HD\,155889 appears quite typical of O-type stars (Fig. \ref{hd155889}). It is mostly soft, but appears slightly harder than that of \hd. The unabsorbed flux in the 1.-2.5\,keV band is a third of the unabsorbed flux in the 0.5-1.\,keV band, whereas it was one tenth for \hd. A faint high-energy tail required 3 temperature components for the best fit (see Table \ref{tab:fit}). With the same assumptions as before, the bolometric luminosity was evaluated to $2.5\times10^{38}$\ergs\ and the \lxlb\ ratio is therefore $-6.68$, again quite typical of O-type stars \citep{naz09}. There is thus no X-ray overluminosity for HD\,155889 and the companion is obviously not the dominant X-ray emitter in the system. In addition, the lightcurves show no significant variations during our observation: the X-ray emission is therefore in reasonable agreement with the expectations of the wind-shock model.

\end{document}